\documentclass[aps,twocolumn]{revtex4}
\usepackage{graphicx}
\usepackage{color}
\usepackage{bm}
\begin{document}
\title{Bias induced ferromagnetism and half-metallicity in graphene nano-ribbons}
\author{Rita Maji}
\affiliation{School of Physical Sciences, 
National Institute of Science Education and Research, HBNI, Jatni - 752050, Odisha, India}
\author{Joydeep Bhattacharjee}
\affiliation{School of Physical Sciences, 
National Institute of Science Education and Research, HBNI, Jatni - 752050, Odisha, India}
\email{jbhattacharjee@niser.ac.in}
\begin{abstract}
Towards spin selective electronics made of three coordinated carbon atoms, here we computationally 
propose robust and reversibly bias driven evolution of pristine undoped graphene nano-ribbons(GNR) 
into ferromagnetic-semiconductor, metal or a half metal, irrespective of their edge configurations.
The evolution is a result of a rare ferromagnetic(FM) order emerging among nearest neighbouring(n-n) sites,
in positively biased regions in their in-homogeneous bias unit-cells,
in attempt to cooperatively minimise on-site Coulomb repulsion and kinetic energy, while maximising
localization of electrons at the positively biased sites.
The phenomenon appears to be a general property of in-homogeneously biased  Coulomb correlated bipartite systems.
Consequences are particularly rich in zigzag edged graphene nano-ribbons(ZGNR) due to the
contest of bias driven n-n FM order and the inter-edge antiferromagnetic order inherent to ZGNRs,
leading to systematic closing of gap for one of the spins, amounting to bias controlled 
unmissable opening of window for FM-semiconducting and half-metallic transport.
\end{abstract}
\maketitle
Sheets, ribbons and tubes made of three coordinated $sp^2$ hybridized carbon(C) atoms
can be semiconducting or metallic \cite{gnr-width,rev-gnf} depending on their shape, size and edge
configuration. 
They have been thus long anticipated to constitute a framework for carbon based 
electronic circuitry at nano-scale\cite{Avouris2007}.
$2p_z$ electrons of these carbon atoms, if rendered unpaired due to lack of coordination, like at defects or edges,
are source of local magnetic moments\cite{rev-magnetic,lieb} and ferrimagnetism in their vicinity.
Tuning magnetism of these electrons to add spin-selectivity to carbon based circuit elements, 
has been an active area of research\cite{Han2014} in the last two decades or so.
A large variety of proposals and demonstrations made in this direction based on 
structural \cite{EXPT-strc,width-ribbon,strain,gnr-structure}, 
physical\cite{irr-expt,irr-DFT,zgn-efield,agn-efield,defects-vacancy,rev-magnetic} and 
chemical\cite{expt-chemf,expt-moldope,sanchez,chem-f,doping,DFT-vacancy,chmf-rev} functionalization, 
have promised the possibility of magnetic semiconductors and half-metal\cite{zgn-efield,hm,hm-edge,hm-tr} 
primarily in zigzag edged graphene segments, ribbons and tubes\cite{expt-edge,expt-zgnr,rev-magnetic}. 
Realization of such proposals into commercially viable devices is challenged by the 
stringent requirement of precise control over their shape, size, and edge configurations.

Based on mean-field and ab-initio computation of spin resolved electronic structure, 
this suggests a plausible alternate, as it implies that in principle 
any ribbon of three coordinate carbon network can be controllably turned into a 
ferromagnetic(FM) semiconductor or metal, and a half-metal, through in-homogeneous biasing,
provided, the localization of electrons in the positively biased regions be
condusive for emergence of nearest neighbor (n-n) FM ordering, which we refer here onwards FM$_{n-n}$.
Through a minimal model system, we argue here that such emergence of FM$_{n-n}$ order is a general property of 
in-homogeneously biased bipartite systems, arising primarily as a means to avoid increase of 
on-site Coulomb repulsion and kinetic energy while maximizing response to the external bias.

\begin{figure*}[t]
\centering
\includegraphics[scale=0.43]{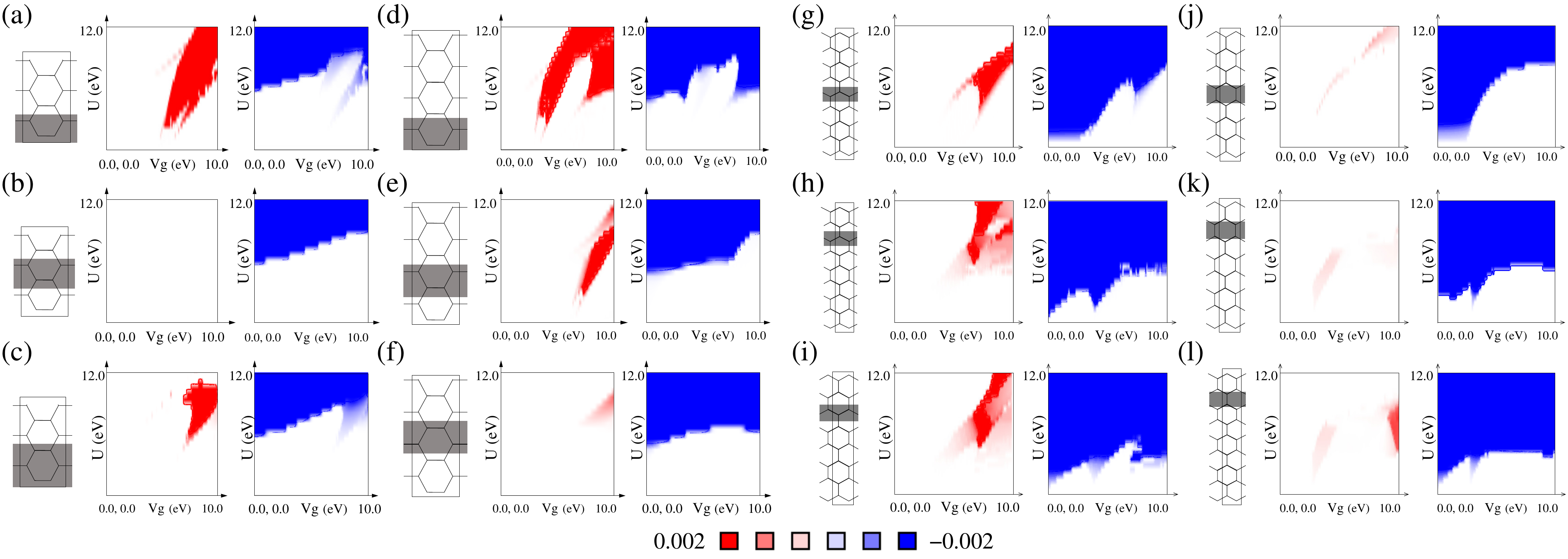}
\caption{spin-correlation(positive \& negative) plot as function of $U$ \& $V^g$ for different bias coverages: 
AGNR(N=6)(a,b,c); AGNR(N=8)(d,e,f); ZGNR(N=18)(g,h,i); ZGNR(N=16)(j,k,l). The degree of influence of width and location
of biased region on the range of $V^g$ and $U$ in which FM$_{n-n}$ would emerge, differs from ZGNRs to AGNRs.
The difference can be attributed to the presence of weak but non-zero n-n FeM order in ZGNRs away from the edges. 
Wider biased region imply weaker localization, and thereby, weak spin separation. These results imply 
significance of localization and inherent magnetic order on emergence of FM$_{n-n}$. }
\label{fig1}
\end{figure*}
In  undoped bipartite systems magnetism is known to arise either at strong coupling\cite{mott-original,mott-bipartite,Fazekas}, 
where the strength of Coulomb correlation is much higher than kinetic energy, leading to n-n anti-ferromagnetic(AFM) order,
or at moderate coupling due to in-equivalence of the two sub-structures\cite{sublattice}, leading to n-n ferrimagntic(FeM) order.
The latter leads to non-zero magnetic moment if the two sub-structures are globally in-equivalent\cite{lieb}.
Majority of proposals referred above are in this category, where the in-equivalence can be due to a host of
reasons, like, vacancy defects \cite{DFT-vacancy}, substitutional doping \cite{doping}, adsorption at sites\cite{adsorption}, 
application of transverse electric field\cite{zgn-efield,agn-efield}.  
On the other hand, FM$_{n-n}$ ordering has been proposed in bipartite systems only upon doping by hole or electrons\cite{Jung2009,Dutta2012}.
Description of magnetism sourced at Coulomb correlation among itinerant electrons,
as derived within the framework of Hubbard model\cite{HUB}
 suggests primarily two classes of mechanisms to 
rationalize FM$_{n-n}$ ordering in bipartite systems upon deviation from half-filling\cite{byndhalf}.
With  $U\rightarrow \infty$, it was shown by Nagaoka \cite{nag} that upon doping by a single hole the
ground state will have FM ordering in attempt to reduce the kinetic energy of the hole, while avoiding occupancy of
a site by more than one electron.
However, Nagaoka-FM has been argued to be not relevant to three coordinated systems \cite{Fazekas}, 
since the loops connecting the n-n sites should not pass through more than four sites for  Nagaoka-FM to sustain.  
In the other class of mechanisms, FM ordering is proposed to be exchange driven, but require a flat or nearly 
flat band \cite{flb1,nearly-flb} at Fermi energy to accommodate electrons emptied from the doubly occupied states without causing 
any or much any increase in kinetic energy.
Itinerant electrons have been also argued\cite{wnr-fb}  to propagate exchange interaction between local moments 
due to flat bands. An approximate meeting ground of the two pathways lead to the Stoner criteria \cite{stonr}, 
which argues that a high $U$ and non-zero 
density of states (DOS) at Fermi energy is necessary for the unequal number of electrons with the two spins 
to be energetically favorable.

To compute spin-polarized electronic structure in the realistic AGNR and ZGNR unit-cells, 
we resort to the mean-field approximation of Hubbard model  within the n-n tight-binding framework:
\begin{equation}
H=-t(\sum_{\langle i,j\rangle,\sigma}c^\dagger_{i,\sigma}c_{j,\sigma}+h.c)+
\sum_{i,\sigma}c^\dagger_{i,\sigma}c_{i,\sigma}(U\langle n_{i,\sigma'}\rangle-V^g_i+V^q_i)
\label{eqn1}
\end{equation}
 $\langle n_{i,\sigma'}\rangle$  being the population of electron   
with spin-$\sigma$ at the $i$-th site due to the occupied states.
Eqn.\ref{eqn1} implies  a self-consistent computation of electronic structure as a function of 
the strength of on-site Coulomb repulsion ($U$) \cite{U-val} between opposite spins, and 
the gate bias $\left\{V^g_i\right\}$, given the lowering of energy due to hopping between 
n-n sites represented by $t= -2.7 eV$ \cite{t-val}.
Coulomb potential ($V^q$) due to electrons at nearest neighboring sites and beyond is calculated 
using the Ewald summation scheme \cite{ewld}. 
We compare our mean-field Hubbard model based results with their counterparts obtained from 
first principles using density functional theory (DFT).
We use a plane-wave based implementation \cite{DFT-QE} of DFT,  wherein, we have used a gradient corrected functional\cite{pbe}
of density to approximately estimate the exchange-correlation contribution to total energy. 

\begin{figure*}[t]
\centering
\includegraphics[scale=0.16]{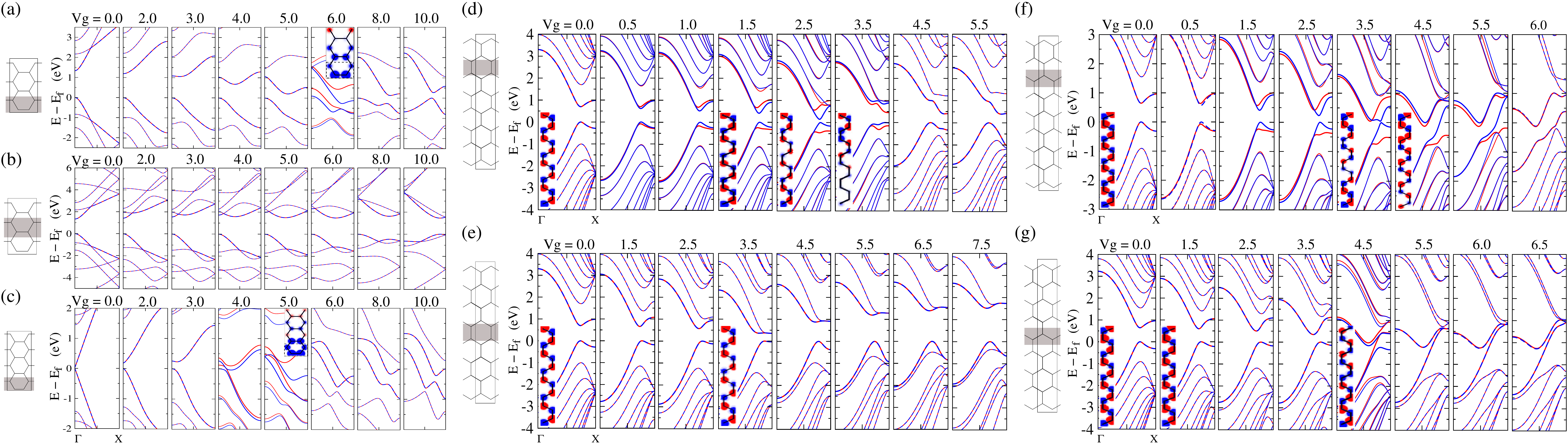}
\caption{ Band structure and spin density (inset) plot based on mean-field Hubbard model for $U$=4.0 eV for different bias coverage:
AGNR(N=6)(a,b); AGNR(N=9)(c);  ZGNR(N=16)(d,e : bias coverage of transverse C-C bonds); 
ZGNR(N=18) (f,g : bias coverage of zig-zag C-C bonds). Emergence of FM semiconducting and half-metallic phases are more
robust and prominent in case of ZGNR due to the contest between emergent FM$_{n-n}$ order and the inter-edge AFM order
inherent to ZGNRs. }
\label{fig2}
\end{figure*}

To probe the nature of magnetic ordering as a function of $V^g$ and $U$, we calculate the average n-n  
spin correlation as:
\begin{equation} 
S=\frac{1}{N_s}\sum^{N_s}_i \frac{1}{\mbox{nn}_i}\sum_j^{\mbox{nn}_i} S_i S_j,
\label{eqn2}
\end{equation}
where $N_s$ is the number of sites per unit-cell, $\mbox{nn}_i$ the number of n-n sites around the $i$-th site, and 
$S_i=\langle n_{i,\sigma}\rangle-\langle n_{i,\sigma'}\rangle$, 
with $\langle n_{i,\sigma(\sigma')}\rangle$  being the population of electron     
with spin-$\sigma(\sigma')$ at the $i$-th site due to the occupied states calculated using the mean-field
approximation of  Hubbard model.
Positive and negative values of $S$ plotted in Fig.\ref{fig1} 
imply existence of FM and AFM or FeM ordering respectively. 
Existence of both thus imply spatial separation between FM and FeM ordering. 


For AGNRs, Fig.\ref{fig1}(a-f) imply rapid consolidation of AFM(FeM) ordering  above $U\sim 2|t|$ with zero(positive) $V^g$.
For $V^g=0$ this is reminiscent of Mott transition \cite{m-itransition} at half-filling ($n=1$) in bipartite lattices.
The trend that with increasing $V^g$ the transition from non-magnetic to the FeM ground state happens with increasing $U$,
is similar to that observed in case of non-magnetic to AFM transition in bipartite lattices with 
increasing deviation from half-filling, and is understood in terms of the additional correlation required
to dominate over the kinetic energy of the excess charges. 
The similarity in trend is expected since with non-zero $V^g$ the biased and unbiased regions both 
deviate locally from half-filling.
With $U>0$ at $v^g=0$, ZGNRs expectedly show n-n FeM ordering and AFM ordering globally between the two substructures.
With increasing $V^g$, quenching of magnetic ordering in ZGNRs below an increasing threshold of $U$ can be understood 
as the dominance of positive bias over on-site Coulomb correlation, leading to occupation of biased sites by
electrons with both the spins.
\begin{figure}[b]
\centering
\includegraphics[scale=0.1]{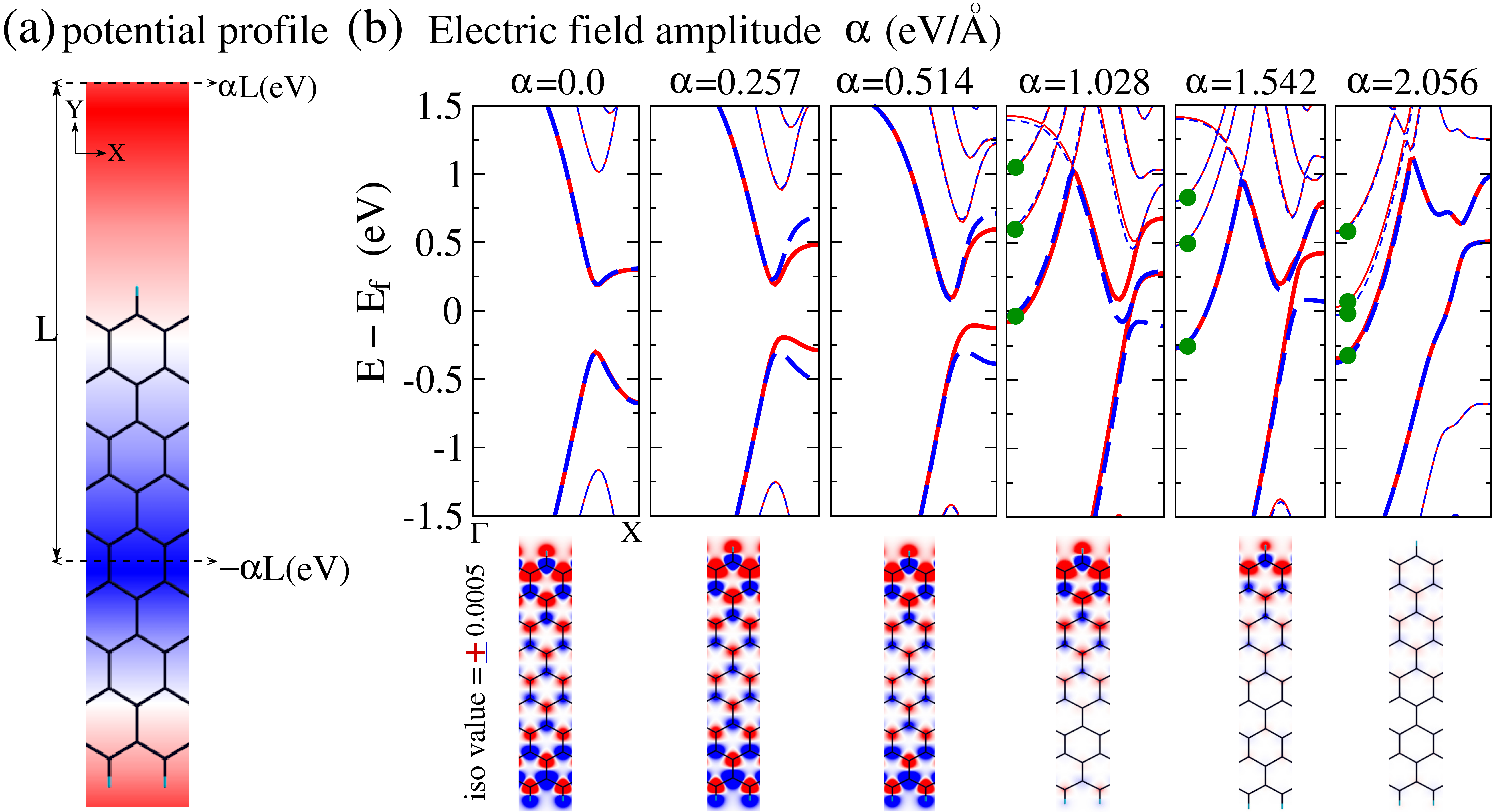}
\caption{ZGNR in presence of sawtooth potential: (a) potential profile ; (b) band structure (above) \& spin density(below).
DFT results are qualitatively similar to their counterparts [Fig.\ref{fig2}] based on mean-field Hubbard model. Bands marked
by the green dots are exclusively due to space charge.}
\label{fig3}
\end{figure}

Emergence of FM$_{n-n}$ ordering is marked by  the positive spin correlation [Fig.\ref{fig1}] and the 
associated lifting of spin degeneracy of the band-structures [Fig.\ref{fig2}] 
over a range of $V^g$ with $U$ moderate and higher. 
FM ordering quenches rapidly in AGNRs [ Fig.\ref{fig1}(a-f)] as the biased region moves away from the edges or are widened.
Similarly in ZGNRs, positive spin correlation is much prominent if the biased region cover zigzag chains of carbon atoms       
parallel to the edges.
Notably, an FM phase of generalized Nagaoka type is known to occur in the mean-field phase diagram of cubic bipartite lattice
\cite{Fazekas} at deviation from half-filling.
Although Nagaoka may not be feasible \cite{Fazekas} in three coordinated bipartite lattices, it 
is beyond the scope of this work to comment on whether it  will be effective 
on the background of increased correlation due to confinement.
However, the trend that the onset of the FM$_{n-n}$ order is more prominent if the biased region is narrow and located closer
to the ribbon edges, clearly suggest that localization of electrons, and consequently the enhanced Coulomb correlation,      
are likely the key associated factors leading to FM$_{n-n}$.
Notably, in case of ZGNRs, if the biased sites cover zigzag(transverse) C-C bonds then the spin at those FM$_{n-n}$ ordered sites
would prefer to be FM(AFM) ordered with sites at both the edges, mediated by the inherent n-n FeM order prevalent outside the biased region. 
Thus in case of biased zigzag sites, mixing of FM$_{n-n}$ ordered state with the dominant spin, and the nearest localized edge state, can    
stabilize both, leading to relative ease in occurrence of FM$_{n-n}$ order compared to that in case of biased transverse sites.   
This is consistent with less positive spin-correlation[Fig.\ref{fig1}(j,k,l)]  in case of biased C-C transverse bonds. 

FM$_{n-n}$ order at the positively biased sites inherently implies lifting of spin-degeneracy   
since it consolidates one of the spins in the biased region. Emergence of the FM-semiconducting,
FM-metallic and half-metallic phases are thus naturally expected as consequences of  FM$_{n-n}$.
However, for AGNRs, onset of FM$_{n-n}$ order is preceded by shrinking of band-gap and 
direct to indirect transition. 
With increasing $V^g$, the bands representing the bonding and anti-bonding orbitals of the biased region 
will come down in energy, while their counterparts associated with the edge farthest from the biased region will have least
change in energies, resulting into a net reduction in band gap, as observed.
Thus the conduction(valence) band edges can be spatially located at the positively(zero) biased regions of the unit-cell.  
In such a scenario a direct to indirect transition is expected, since the conduction and valence band edges,
which are at different $V^g$, tend to have same total energies as the band-gap shrinks, implying that their kinetic energies 
must be different. With increasing $V^g$, further lowering of the conduction band edge leads to complete closure
of the indirect band gap.
The AGNRs are thus most likely to be normal or FM-metal upon emergence of FM$_{n-n}$ order [Fig.\ref{fig2}(a,c)],
although a small window for half-metallic transport can open up serendipitously.
As the biased region moves towards the bulk(middle) of ribbons, lowering of kinetic energy increasingly 
compensates Coulomb correlation, leading largely to non-magnetic ground state and spin-degenerate band structure [Fig.\ref{fig2}(b)] . 

\begin{figure}[t]
\centering
\includegraphics[scale=0.19]{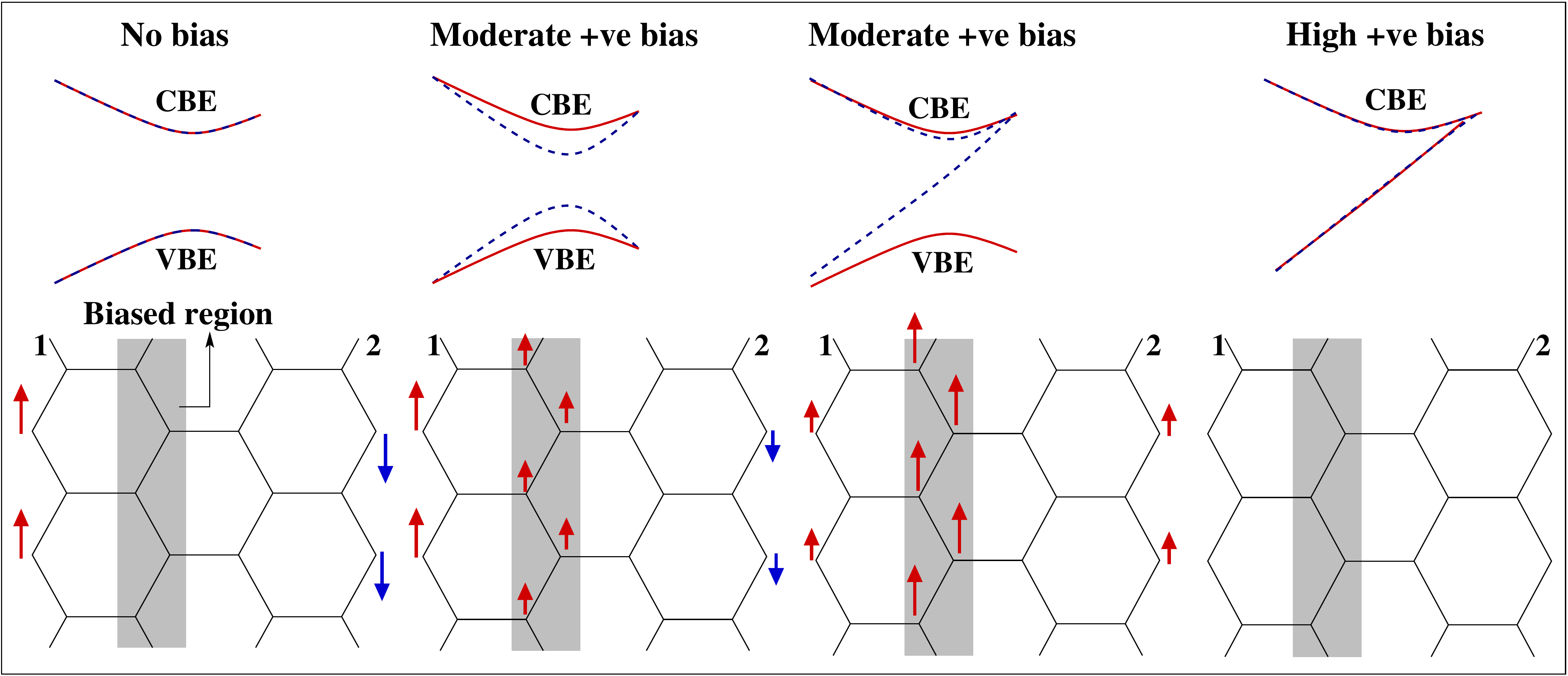}
\caption{Schematic view of band structure (above) \& spin density ($S_i$) (below). 
The edges and the biased regions are mediated by the 
n-n FeM ordering, owing to which, the FM$_{n-n}$ ordering with spin-1({\color{red}{ $\uparrow$}}) in the biased region induces
preference for the same spin at the edges, leading to  suppression of the edge state at edge-2
with spin-2 ({\color{blue}{ $\downarrow$}}). The evolution of the conduction and valence band edges (CBE,VBE respectively)
have been argued in analogy with Fig.\ref{fig2}(f).}
\label{fig4}
\end{figure}

In ZGNRs[Fig.\ref{fig2}(d-g)] the effect of FM$_{n-n}$ ordering is particularly rich, since the FM ordering contests  the 
inherent AFM ordering of the edge states by favoring same spin at the two edges, similar(opposite) to that of the 
FM$_{n-n}$ order if the biased region covers C-C zigzag(transverse) bonds. 
As shown schematically in Fig.\ref{fig4}, the contest supports the edge-states with spin-1 at edge-1,
but suppresses the edge-state with spin-2 at edge-2. 
The consequent systematic closure of gap[Fig.\ref{fig2}(f,g)] exclusively for spin-2,[Fig.\ref{fig4}]
can be understood as a result of effective increase in on-site energy for electrons with spin-2 at edge-2 
due to accumulation of electrons with spin-1 at that edge on account of FM$_{n-n}$. 
Expedient to recall, equal probability  of finding an electron at both the edges, implying same effective 
on-site term at both the edges, leads to closure of band-gap,
as happens for both the spins in the absence of Coulomb correlation.
Indeed, the reduction in n-n FeM order near edge-2 due to the competing magnetic orders,
implies reduction of the effective Coulomb correlation near edge-2 where spin-2 dominates.
Similar effective reduction of Coulomb correlation should happens for both the spins as the 
biased region shifts towards the bulk, since it would increasingly lead to non-magnetic ground state
and spin degenerate band-structure [Fig.\ref{fig2}(e)] , akin to AGNRs. 
Nevertheless, with biased region covering C-C zigzag bonds[Fig.\ref{fig2}(f)], the inter-edge AFM ordering 
clearly evolves into inter-edge FM ordering, leading to evolution of the valence band edge for spin-2 
into a partially occupied dispersive band, which offers a robust window for half-metallic transport.
This reiterating that bias coverage of zigzag C atoms is more effective than that of biased transverse C atoms in 
emergence of FM$_{n-n}$ order.
However, the systematic reduction of band-gap for one of the spins, as argued above, happens in ZGNRs irrespective of  
whether the biased region covers C-C zigzag or transverse bonds, paving the way for robut bias controlled
FM-semiconducting and half-metallic transport.
Thus the same is offered by more realistic wider biased regions as well, at moderate $U$ and low $V^g$, 
which might be technologically relevant. 

Band-structure calculated from first principles using density functional theory (DFT)
with sawtooth potential \cite{DFT-QE} applied in the transverse direction[Fig.\ref{fig3}(a)] of ZGNR to resemble  biased region akin 
to those considered in Fig.\ref{fig1}, has similar trend as [Fig.\ref{fig2}(f)], marked by lifting of
 spin-degeneracy and reduction in band-gap for one of the spins[Fig.\ref{fig3}] within a range of ramp potential. 
This qualitative agreement between mean-field and DFT results is an important validation of the former, 
which only incorporates on-site Coulomb correlation.  
Furthermore, the reported agreement of DMRG, QMC and ED results\cite{wnr-fb,MF-comparison} with DFT in rationalizing 
FM ground state in doped AGNRs at moderate $U$, implies the confidence that our mean-field results will also be valid 
with improved consideration of correlation.
Structural relaxation \cite{method-bfgs}using forces derived from DFT ground states appear to indicate that the ribbons should maintain their 
structure intact in the range of ramp potential for which the FM$_{n-n}$ order exists. 
However, in addition to the FM$_{n-n}$ ordering, we also find highly dispersive free electron like
bands around Fermi energy[Fig.\ref{fig3}(b)], whose origin is traced to accumulation of space charge between the periodic 
images of the ribbon. Upon emergence of FM$_{n-n}$ order in the ribbon, the space charge also acquires a net non zero magnetic moment.  

\begin{figure}[t]
\centering
\includegraphics[scale=0.1]{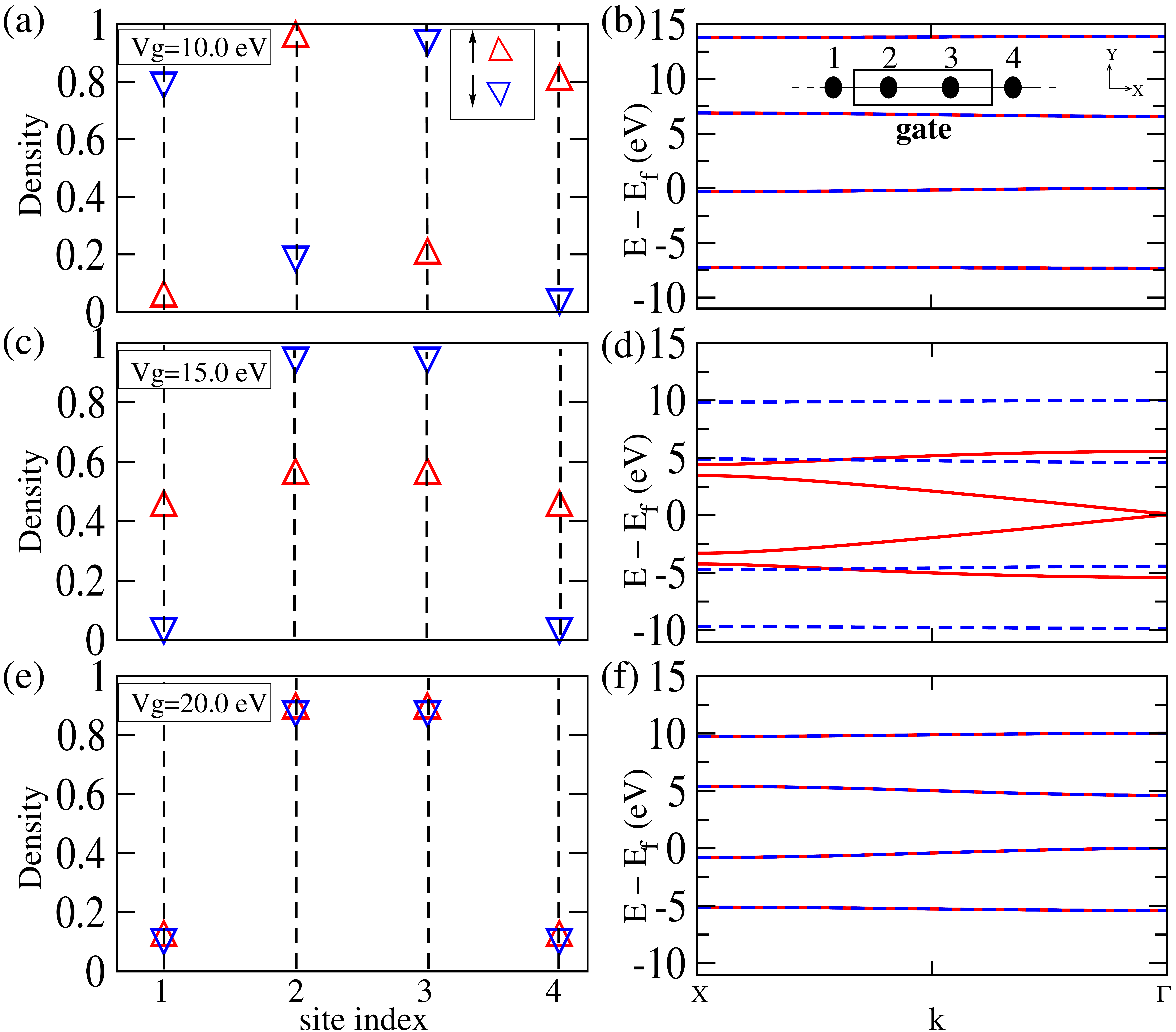}
\caption{Density(up,down) and band structure of 4-site linear chain from mean-field Hubbard for $U$=15.0 :$V^g$=10.0 eV (a,b); 
$V^g$=15.0 eV (c,d); $V^g$=20.0 eV (e,f). The evolution of band structure is qualitatively similar to those in Fig.\ref{fig2}, implying
generality of the FM$_{n-n}$ order.}
\label{fig5}
\end{figure}

\begin{figure}[t]
\centering
\includegraphics[scale=0.08]{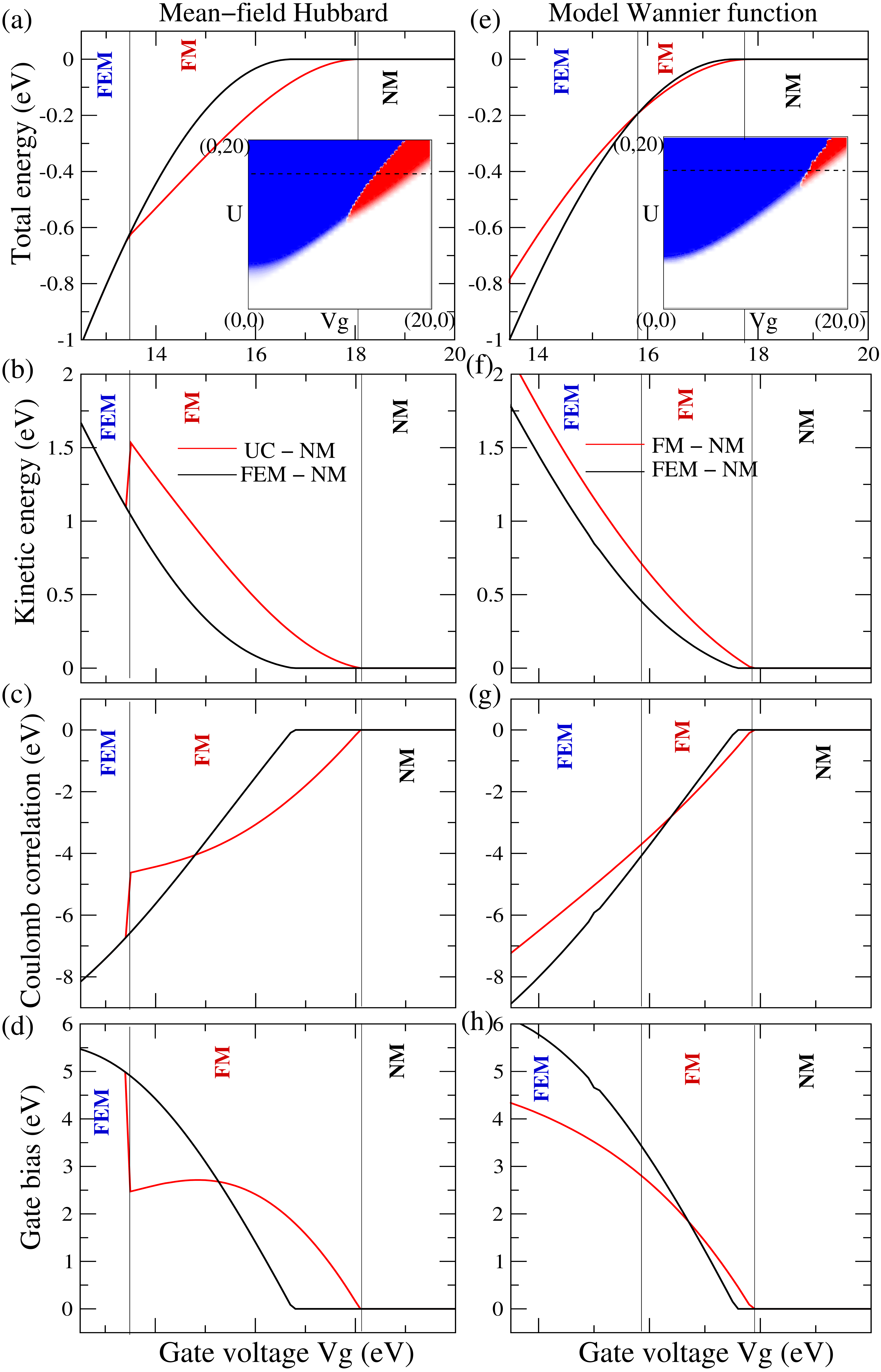}
\caption{Biased site spin-correlation plot (inset) and different energy contributions for $U$=15.0 from : 
mean-field Hubbard model(a,b,c,d) \& analytic WF based calculation(e,f,g,h).
The qualitative similarity of the energetics responsible for
emergence of the FM$_{n-n}$ order, imply that the mean-field results are valid beyond the mean-field approximation. }
\label{fig6}
\end{figure}

To check the generality of our results beyond the three coordinated networks we resort to a minimal unit-cell, 
which can exhibit the FM$_{n-n}$ order if it exists. We choose a unit-cell of four consecutive sites,
of which two neighboring sites are biased[inset,Fig.\ref{fig5}(b)]. 
Spin-correlation between the two biased sites[inset, Fig.\ref{fig6}]
as a function of $V^g$ and $U$ has similar generic features as in Fig.\ref{fig1}.
thus hinting at the generality of the n-n FM order as a property of non-uniformly biased bipartite systems. 
Notably, if we do not consider non-zero crystal momentum, then the positive spin-correlation does not exist,
although the trend of spin-correlation as a function of $V^g$ and $U$ obtained 
using the mean-field Hubbard model, agrees qualitatively with that obtained using exact-diagonalization(ED).
Itinerant electrons described by dispersive bands at Fermi energy are thus likely to play important role in 
manifestation of FM$_{n-n}$ order, as suggested by Fig.\ref{fig2}.

The evolution of spin polarized charge densities [Fig.\ref{fig5}(a,c,e)] with increasing $V^g$ suggests spin-separation
between biased and unbiased region similar to that observed in ribbons.
Such a spin-separation can possibly be rationalized by noting that increased occupancy of
the biased sites by electrons with both the spins, upon increased $V^g$, would in turn increase potential energy
due to on-site Coulomb repulsion. Thus to keep potential energy low, each of the biased sites will prefer to be
dominated by electrons with one of the two spins. However if two such neighboring sites have opposite spins, 
as would be in a FeM ground state, then the wave-functions for both the spins will have rapid variation from site to site, 
leading to high kinetic energy.
Instead, if wave-functions of one type of spin spans the biased sites more than those with the other spin, as evident from
the charge densities [Fig.\ref{fig5}(a,c,e)], then the wave-functions can be smoother than their FeM counterparts, implying lesser kinetic energy
while allowing lower on-site Coulomb repulsion as well. 
The evolution of charge densities[Fig.\ref{fig5}(a,c,e)] clearly implies electrons to be more itinerant for one spin 
than for the other upon emergence of  FM$_{n-n}$, leading to lifting of spin degeneracy [Fig.\ref{fig5}(c,d)] akin to that in [Fig.\ref{fig2}(a,c)]. 
Upon further increase of $V^g$[Fig.\ref{fig6}(b,c,d)]
the lowering of potential energy due to $V^g$ dominates over the increase of potential energy due to on-site
Coulomb repulsion, leading to occupancy of bias sites by both spins, resulting into return of non-magnetic ground state and spin degenerate
band-structure[Fig.\ref{fig5}(e,f)].
Thus with higher $U$ a higher $V^g$ is required for the FM$_{n-n}$ order to quench, which is also consistent
with the trend observed in Fig.\ref{fig1}.

To quantitatively justify the  mechanism anticipated above, we partition the total energy[Fig.\ref{fig6}(a)] of the  
 unconstrained(UC) ground state into kinetic energy and potential energies due to on-site Coulomb repulsion 
and applied bias, and compare those with their counterparts in a non-magnetic(NM) and FeM ground states.
NM ground state is obtained by assigning same charge density for both the spins. For the FeM ground state, 
charge density for one of the spins is assigned to be the mirror image of that for the other spin about the centre of the unit-cell.
We choose the energetics of the NM ground state to be the reference.
Fig.\ref{fig6}(c,d) suggests that the  UC ground state with FM$_{n-n}$ order has lower energy than NM and FeM ground states,
owing initially to lowering of potential energy due to gate bias, but subsequently and primarily due to lowering of 
on-site Coulomb repulsion facilitated by separation of spin between positively biased and unbiased sites.  
Notably, although the FeM ground state has a lower kinetic energy than the UC ground state, the degree of 
localization at the positively biased site offered by the former is much lower than that due to the latter.    
Thus an FeM ordered state with the same degree of localization as offered by the UC ground state must have higher kinetic
energy than the latter. Thus the observed FM$_{n-n}$ order is a result of attempts to minimize on-site Coulomb repulsion
and kinetic energy in conjunction with each other, while maximizing localization of electrons at the positively biases
sites.


In addition to agreement with DFT results, to demonstrate further that the observed FM$_{n-n}$ order 
is not limited by the mean-field approximation,
we resort to simple analytical models for Wannier functions(WF), to represent the ground state of the
four site unit-cell at half-filling.
WFs are linear combination of wave-functions and can be chosen to be real and localized largely within a unit-cell
in one dimension. Thus in place of $2N_k$ wave-functions for each spin, $N_k$ being the number of allowed
wave-vectors in the first Brillouin zone, we can choose two WFs for each spin to represent four electrons.  
We approximate WFs to be non-zero only within a unit-cell.
Such approximate WFs describing the non-magnetic(NM) ground state can be of the following general form:
\begin{eqnarray*}
 \phi^{NM}_{1,\uparrow/\downarrow}&=&\left(a,b,c,d\right),\nonumber\\
 \phi^{NM}_{2,\uparrow/\downarrow}&=&\left(e+\frac{c.f}{a},g+\frac{d.f}{b},-f-\frac{a.e}{c},-f-\frac{g.b}{d}\right),\nonumber
\end{eqnarray*}
which are constrained to be orthogonal to each other.
Similarly, two orthogonal WFs to describe the FeM ground state can be approximated as:
\begin{eqnarray*}
 \phi^{FeM}_{1,\uparrow}&=&\left(a,b,c,d\right),\nonumber\\
 \phi^{FeM}_{2,\uparrow}&=&\left(e+\frac{c.f}{a},g+\frac{d.f}{b},-f-\frac{a.e}{c},-f-\frac{g.b}{d}\right),\nonumber\\
 \phi^{FeM}_{1,\downarrow}&=&\left(d,c,b,a\right),\nonumber\\
 \phi^{FeM}_{2,\downarrow}&=&\left(f+\frac{g.b}{d},f+\frac{a.e}{c},-g-\frac{d.f}{b},-e-\frac{c.f}{a}\right),\nonumber
\end{eqnarray*}
where the $|\phi_{i,\uparrow}|^2$ is mirror image of  $|\phi_{i,\downarrow}|^2$ with respect to the middle of the unit-cell.
Finally, orthogonal WFs with FM$_{n-n}$ order can be approximated as:
\begin{eqnarray*}
 \phi^{FM}_{1,\uparrow}&=&\left(a,b,b,a\right),\nonumber\\
 \phi^{FM}_{2\uparrow}&=&\left(c,d,-d,-c\right),\nonumber\\
 \phi^{FM}_{1\downarrow}&=&\left(e,f,f,e\right),\nonumber\\
 \phi^{FM}_{2\downarrow}&=&\left(g,h,-h,-g\right).\nonumber
\end{eqnarray*}
The number of independent variables chosen to express the WFs are determined by the  symmetry of the 
spin densities [Fig.\ref{fig5}(a,c,e)] and orthogonality of the states.
Total energies of ground states are calculated in the basis of the approximate WFs within the 
Hubbard model without mean-field approximation.
For each class of WFs, ground state is obtained by finding the global minima of total energy
using the cylindrical algebraic decomposition algorithm \cite{mathwlf}.
Kinetic energy and potential energies due to Coulomb repulsion and external bias are estimated 
using the WFs corresponding to the ground states. Notably, Fig.\ref{fig6}(e-h) implies emergence of FM$_{n-n}$ order
in exactly the same pathway as suggested within the mean-field approximation of Hubbard model[Fig.\ref{fig6}(a-d)].
These agreements  indeed confirms the FM$_{n-n}$ order
to be a general property to be computationally observed beyond mean-field approximation, 
in in-homogeneously biased bipartite systems. 

To summarize, in this work we computationally propose emergence of localization induced ferromagnetic(FM) order among 
nearest neighboring (n-n) sites in positively bised regions of non-uniformly biased bipartite systems, 
as a generic outcome of efforts to minimize Coulomb repulsion with minimal loss of itinerancy
of electrons, while maximizing localization of electrons in the positively biased sites.
The n-n FM order is computationally demonstrated here to  exist in ribbons of three coordinated carbon networks
irrespective of their edge configurations, as well as in a minimal model,
where the associated lifting of spin-degeneracy leads to metallic (normal, ferromagnetic and half) phases.
In armchair edged semiconducting ribbons the metallic phases are presided by direct to indirect transition,
while in zigzag edged graphene ribbons, their inherent inter-edge AFM order contests the bias driven FM$_{n-n}$ order,
leading to systematic closer of gap for one of the spins resulting into robust window for 
FM semiconducting and half-metallic transport.
These results are expected to encourage a conceptually new pathway for voltage controlled opening of 
windows for half-metallic transport in two dimensional systems in general. 
\section{Acknowledgements}
Calculations have been performed in a high performance computing facility funded by 
Nanomission(SR/NM/NS-1026/2011) of the Dept. of Sci. Tech. of the Govt. of India. RM acknowledges
financial support from the Dept. of Atomic Energy of the Govt. of India.

\end{document}